# Could a Computer Architect Understand our Brain?

**Valentin Puente-Varona**

University of Cantabria, 39005 Santander, Spain
* Correspondence should be addressed to V.P.V. (vpuente@unican.es)

**This paper presents a highly speculative model encompassing the cortex, thalamus, and hippocampus of the mammalian brain. While the majority of computational neuroscience models are founded upon empirical evidence, this model is predicated upon a hardware proposal for a machine learning accelerator. Such a device was designed to perform a specific task, such as speech recognition. The design process employed the principles and techniques typically used by computer architects in the design of devices such as processors. However, it also sought to maintain plausibility with biological systems in accordance with the current understanding of the mammalian brain. In the course of our research, we have identified a functional framework that may help to fill the gaps in current neuroscience, thereby facilitating the explanations for many elusive cognitive-level effects. This paper does not describe the device itself or the rationale behind the design decision, but instead, it presents a concise description of the derived model. In brief, the model provides a functional definition of the cortical column and its structural definition by the minicolumns. It also offers a descriptive model for the corticothalamic and corticostriatal loops, a functional proposal for the hippocampal complex, and a simplified view of the brainstem circuitry involved in auditory processing. The proposed model appears to provide an explanation for a number of cognitive phenomena, including some ERP effects, bottom-up and top-down attention, and the relationship between phenomena such as the cocktail party effect, anterograde and retrograde amnesia following hippocampal complex damage, and so forth.**

## Introduction

A complex system is defined as a system comprising a multitude of parts that interact in a non-simple manner (1). Such systems may be biological or artificial. In light of this definition, computers may be the most complex man-made system. To cope with such complexity, as (1) argue, hierarchical layering and abstraction are the central scheme used. Each expert deals with one layer of the hierarchy, exposing only an interface that abstracts away the implementation details to the users above. Computer architects are responsible for facilitating the transfer of technological innovations from the lower levels of the computer hierarchy to the upper levels of the hierarchy, specifically the software. The outcome of technologists' efforts is the creation of resources (e.g., transistors exposed to the architect as registers, functional units, classes of memory, etc.). Computer architects are responsible for transforming these resources into performance, efficiency, resilience, etc. The many software layers above (management engines, hypervisors, operating systems, applications, etc.) can exploit such capabilities. A comparable methodology is necessary to reverse-engineer any of the layers, for instance, the processor.

The current neuroscience methods appear to be inadequate even for the reverse engineering of a processor designed in 1975 (2). This may indicate a lack of consideration for the existence of layer diversity at the base of the hierarchy. Could processor designer mindset be useful to reverse-engineer a brain? As in computer systems, if we assume what (1) postulates, the brain will be architected by evolution around a hierarchy. From this perspective, it seems reasonable to hypothesize the existence of such organization and to choose the "right layer" to break the code. For instance, attempting to reverse-engineer by brute force a contemporary processor (with tens of billions of transistors, as opposed to the mere 3510 transistors of the MOS 6502(2)) appears to be an unattainable goal. From the perspective of a computer architect, it seems impossible to comprehend the intricacies of components such as cache coherence controllers at the transistor level. Similarly, attempting to comprehend the operation of the processor by replicating the response of hundreds of millions of lines of code from a contemporary software stack to user input appears to be an inefficient use of resources. In our view, this extreme case may be analogous to the current deep learning revolution, which, while useful now, seems destined to fail in the medium term(3).

When reverse engineering a system, the question is how to select the "right" layer to understand the "core" of the system. Regarding the case of the processor, the majority of contemporary system hardware complexity can be attributed to mechanisms required to circumvent physical constraints, such as those related to memory timing or bandwidth (4). Addressing these requires a complex memory hierarchy, out-of-order execution, multiple speculation engines, etc. It is likely that the fragility of the biological substrate imposes constraints on a scale that is orders of magnitude greater. Consequently, the difficulties faced by conventional approaches in untangling the hierarchy complexity will be orders of magnitude harder than those faced in the case of the most complex processor. Despite the complexity of the computer, the program-stored concept (5), which is the fundamental element of the system's architecture, is remarkably simple. It can be understood as a foundation upon which to progressively unravel all the intricacies of the layers above and below. A reasonable approach to solving this issue would be to select a complex but highly specific task where the computer would be expected to excel. This could then be used as a basis for designing a system that, obeying the known physical substrate constraints, can mimic it. It is important to consider that the design should be capable of being extended to other uses, that is, generalizable. The fundamental tools are to have an intuitive grasp of the system and the capability to simulate its behavior in the "right" detail.

In (6) we employ this approach to develop a highly speculative algorithmic construct of the mammalian neocortex and certain key support structures. Inspired by cortical regularity (7,8), and following the footsteps of (9), we consider the possibility of designing a system capable of learning, predicting and identifying raw human speech. The final design was co-inspired by some of the current neuroscience knowledge and commonly used techniques in computer architecture such as speculation and caches. Bearing in mind biological plausibility, was critical to resolve some of the issues found in the design process.

The model obtained is not only able to solve the challenge, but also to provide some bridges that may connect neuroscience with cognitive level phenomena (e.g., Mcgurk and cocktail effects) or answers to longstanding questions such as the relevance of the electroencephalogram, the effects of astrocytes on learning, the mechanisms behind event segmentation, etc. Next, we succinctly describe the findings as a model for the latter to present its performance and a more detailed view of the phenomena that can be explained.

## Architecture of the Cortical Column
### Pyramidal Cell form Sequence Memories

Pyramidal cells are the foundation of the model. We hypothesize them to be sequence predictors. They work together in small groups to form a Sequence Memory. The spike produced in the distal dendritic arbor serves as a coincidence detector. When a cell generates an EPSP, the immediately preceding dendritic spikes serve to predict which cell in the group will fire next. The cells that fire in the group (~2%) do so in a synchronous manner. The group of cells that fire represents a code in a temporal sequence that arrives as input through the proximal arbors. The dendritic branches where the dendritic spikes have occurred previously, represent the context of the value. The number of different input values that a sequence memory can receive is small (4~8). From the fundamental principles of combinatorics, it can be demonstrated that the greater the number of branches in a pyramidal cell, the greater the number of potential contexts for an input code. Consequently, the

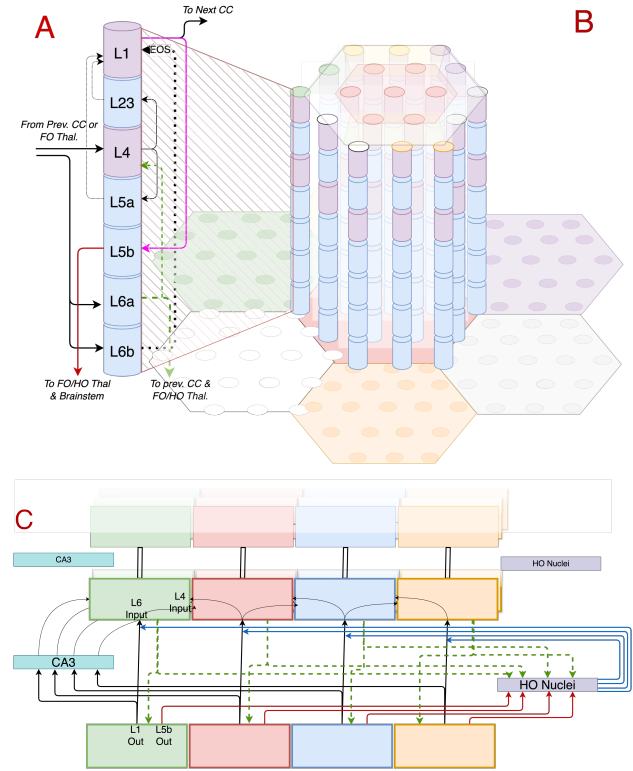

**Figure 1: Architecture of the cortical column (A)** Minicolumn architecture and functional connectivity depicted in the text. **(B)** Integration of the cortical column in upper levels of the hierarchy, including lateral contextualization for a bidimensional cortex, **(C)** Sketch of the relation between two levels in the cortex architecture

Sequence Memory may be capable of tracking an exceptionally large number of sequences. The morphological properties of the distal arbor determine the storage capacity of the Sequence Memory. L23 and L5a pyramidal cells in the cortical column work in this way. The learning process involves repeating in any order the input sequences. Sequence Memories excels on learning multiple sequences concurrently.

### Short-term Learning Modulation

The Sequence Memories learns according to the novelty of the data. We hypothesize that astrocytes control this by regulating the extracellular calcium ($[Ca^{2+}]_0$) in the synaptic cleft. The novelty of the data is determined by averaging the prediction accuracy of the most recent inputs (e.g., by using an exponential moving average). If cell activations were predicted for a given time, there will be an influx of most $[Ca^{2+}]_0$ from the astrocytic network in all the synaptic clefts of the Sequence Memory. Otherwise, there will be an efflux. The relationship between prediction accuracy and $[Ca^{2+}]_0$ follows a logistic function, where the steepness of the function is determined by the cholinergic system. The Sequence Memory has two states, corresponding to the degree of knowledge of the input: *known* and *unknown*. Different thresholds are employed to transition between the two states. In each state, the logistic function is adjusted to its threshold, resulting in a hysteresis cycle. This cycle serves to prevent the system from oscillating between the two states.

### Input Value Projection and Replication

To constrain the diversity of inputs to the Sequence Memory, a projection is made via dense inhibitory networks. The task of such a network is to map the diversity of the raw input onto a small number (4~8) of clusters. To achieve that, the input is densely wired but with very few synapses connected (e.g. 70% existing synapses, with only 10% connected). As the system is exposed to actual information, unused synapses in active dendritic segments are removed via heterosynaptic plasticity. Over the course of the system's lifespan, non-active dendritic segments are gradually pruned.

In the cortical column the projection of L23 and L5a is carried out by L4. In order to maintain the capacity for disambiguation, the system uses a strategy of intensive replication. Since L4 projections depends on the initial connectivity, which is stochastic, having multiple replicas of L4+L23/L5a is necessary. The number of replicas depends on the input traits that are subordinate to the evolutionarily defined traits. It can be on the order of tens or a few hundred in an unreliable and noisy substrate like the biological.

### Sequence Segmentation and Cortical Column Output

L6b is a Sequence Memory that is dedicated to the division of the input sequence of the cortical column. While it shares similarities with L23/L5a, it possesses two distinctive properties: it perceives raw input (i.e., not a projection) and, if the proportion of input is unexpectedly high (i.e., most of active cells were not predicted), it produces an End-of-Sequence (EOS) at the cortical column level. EOS initiates in L1 a projection process analogous to that performed by L4. In this instance, the inputs are the distal branches predicted in the preceding time step by L23 and L5a. Then, at EOS, L1 generates an output value for the cortical column. Since dendritic spikes are context dependent, such a value will represent the sequence identified by L23/L5a. L6b has a short temporal constant (i.e., the knowledge state is evaluated considering the most immediate past). This approach allows the construction of brief sequences (approximately four values) that can benefit from the hierarchical organization, thereby reducing the synaptic load.

### Backward Prediction

The functional target of L6a, despite receiving the same input as L6b, is a Sequence Memory of a different nature. It has a markedly larger temporal constant and it will produce next value predictions for previous cortical columns on the hierarchy and for the thalamus. These predictions are modulatory and are primarily employed to modulate the inhibition results of projectors (L1 of the previous cortical columns and L4 of the current column) and to filter forward sequence predictions. Additionally, the L6a knowledge state defines the knowledge status of the cortical column, which is fundamental to controlling the long-term learning modulation.

### Forward Prediction

The final Sequence Memory in the cortical column, L5b, is utilized to predict the subsequent outputs from the cortical column. To achieve this, when an EOS is reached, L5b learns the L1 sequence of outputs. Once the cortical column has learned its input, L5b will provide a prediction for the next value. This prediction represents the next segment of the sequence that is to come. This value is transmitted to the relay cells in the high-order thalamic nuclei and compared with the next cortical column's backward predictions. If a match is found, the prediction is forwarded, thereby reducing the time required to identify the sensory stream at a higher level in the hierarchy. If such agreement is found, the prediction is recurrently fed into L5b, which allows for the prediction of multiple segments in the future.

### Cortical Minicolumn

The sizes of all sequence memory are similar, the replication observed in L4 should be present throughout the whole cortical column. Hence all previously cited layers are structured in the same way, and anatomically perceived as minicolumns (7). To improve EOS stability, it is necessary to increase the size of L6b (combine more than one minicolumn input). To enhance fault tolerance, L1 cells are also connected to multiple minicolumns. It is likely that this will be achieved by maintaining a similar cell count per functional group, which will result in some spatial overlap at these layers.

## Function of the Hippocampal Complex
### Long-term Learning Modulation and Gating

In contrast to the prevailing models of the hippocampus, such as those proposed by (10,11), its functional goal is to assist the cortex during learning transients. In the event that the cortical column is in *unknown* state (i.e., L6a is in *unknown* state), the cortical column input is rerouted to a specific area of the CA3 in the hippocampus. CA3 is a Sequence Memory that, until it reaches the *known* state, prevents the entire cortical column from learning. When CA3 reaches the *known* state, via the cholinergic system, enhances learning in all the sequence memories within the corresponding cortical column that requested its assistance. This is achieved by modifying the shape of the logistic function utilized in short-term modulation and increasing the permanence of changes produced by STDP. The remaining portions of the hippocampus (CA1/DG/etc.) are responsible for the dynamic allocation of sections of CA3. Consequently, in contrast to the prevailing paradigm, the grid, location, direction, and other spatial variables are merely epiphenomenal manifestations of this transient phenomenon. The cells in question do not represent anything in the environment; rather, they are indicative of the location within the cortex of the cortical column in a distressed situation.

Once the cortical column has reached a known state, the resources allocated inCA3 are freed (i.e., as a cache in a processor, CA3 areas are shared by multiple regions in the cortex) and the learning boost is disabled. The deallocation of CA3 effectively disables learning in L4, resulting in the freezing of its projections. L1 learning is disabled if there is no cortical column affected in *unknown* state (i.e., no CA3 allocated above). To prevent the learning of noise due to the short temporal constants, L6b learning is disabled if there is no CA3 allocated below.

### Sharp-Wave Ripples
The CA3 predictions are recurrently used to facilitate the convergence of L4. When the CA3 is in a known state, the predictions are utilized recurrently, and an efferent copy sent to the L4 layer of the cortical column in distress. This process will be completed at a rapid pace, in accordance with the theta rhythm. It is anticipated that the CA3 predictions will be wider than the expected 2% due to the correspondence with multiple values from near sequences. This approach will facilitate the convergence of the L4 inhibition process towards a stable value in a more expedient manner. This is necessary because if the convergence of L4 is too slow, it will result in the transmission of more spurious values to L23/L5a. SWR addresses the issue of L4 being unable to increase the learning rate above a certain value in the regular regime due to physical and functional limitations. This reduces the synaptic load of the cortical column, allowing for the faster accommodation of incoming knowledge.

## Architecture of the Cortex
### Lateral Contextualization
Given that sensory cortices receive input distributed topographically, it is to be expected that as one ascends the cortical hierarchy, this input should be laterally distributed to adjacent cortical columns. This should be done in a progressive manner, and most likely, regularly through all the levels. The cortical column employs a mechanism, denoted as lateral contextualization, which functions as follows. The output of a cortical column is sent to the L6a/b of the aligned cortical column in the upper level. The L4s in the central minicolumns are connected in a similar manner. In contrast, the L4 in the periphery of the next cortical column input originates from a neighboring cortical column at a lower level.

Consequently, the formation of sub-sequences (i.e., temporal reference frames) is contingent upon the aligned cortical column in the preceding level, while the output value is contingent upon both aligned and neighboring cortical columns. This process enables the effect of each sensor to propagate across numerous cortical columns in the upper levels of the hierarchy. The top levels of the hierarchy lack lateral contextualization, thus enabling the operation of top-down attention mechanisms.

### Reinterpretation of Feedback as Inhibition Stabilization
In contrast to the prevailing view (12), backward predictions from the cortical column are not utilized as a sort of error signal. It is a modulatory input for the inhibition process of L1 and L4. Inhibition processes, which are likely based on k-WTA, are unreliable in the fragile biological environment, especially during learning. The next cortical column prediction will assist L1 in attaining a stable value. This same prediction is employed in L4 of the same cortical column, where it is generated for the same reason.

L6a is also employed to filter sequence prediction from the preceding column in the thalamic high-order nuclei. Consequently, a disruption during a chain of prediction will result in an increase in activity in the originating cortical column. Such a discrepancy between L5b and the subsequent L6a is not an error that necessitates correction; rather, it is an indication that the supragranular layers in the source cortical column should be reactivated, as the sequence cannot be maintained through prediction.

### Cortical Expansion
As we move up the hierarchy to alleviate the low symbol diversity, increase the fault tolerance of the system, and increase the capacity of the cortex, each cortical column output is sent to multiple subsequent cortical columns. This allows a higher regularity across levels since each cortical column architecture will be the same. Only the connectivity between cortical columns is predetermined. Such replication is consistent with the metabotropic nature of L6a axons, as there will be multiple predictions available to both modulate L1 output and filter forward predictions. This also seems consistent with the strikingly homogeneous anatomical appearance of the cortex (13).

## Architecture of the Thalamus
### Attention Mechanism
Lateral contextualization prohibits column-to-column matching of forward predictions. Allowing them may violate input order in supragranular layers of off-centered minicolumns. This will disrupt cortical symbol output. The high-order nuclei of the thalamus are responsible for comparing the ascending (i.e. L5b) and descending (i.e., L6a) predictions of the region where the lateral contextualization occurs. When there is a match, forwarding is enabled. This forms the bottom-up attentional mechanism of the cortex.

A complementary case is when at the top of the hierarchy, where no lateral contextualization is used, forwarding can occur without the intervention of the corticothalamic loop. When cortical column predictions are accepted, supragranular layers are deactivated to conserve energy by inhibiting L4. At this point lateral contextualization synchronized input is not required, allowing HO thalamic nuclei to accept out-of-sync predictions. L5b output is controlled by the attentional mechanism. Thus, L5b will forward predictions to the thalamus

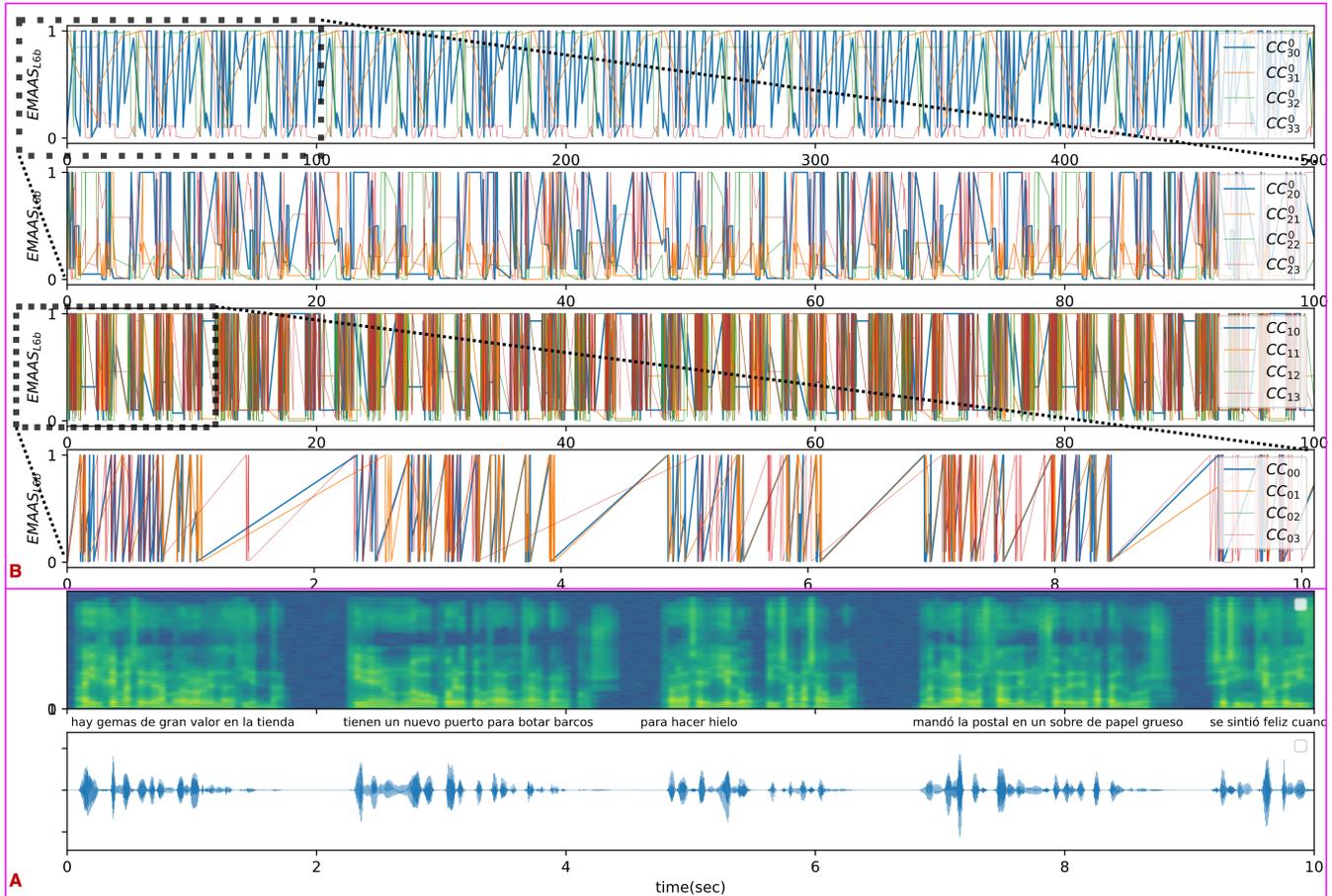

**Figure 2: Synthetic Auditory Cortex response. (A)** Raw signal and spectrogram of a paragraph of ten sentences replayed reiteratively, **(B)** Cortical response from Core, Belt, Parabelt, and TPO of a 4x4 cortex with 24 inputs. L6b EOS are presented, which are the moving average of the anomaly score of the instantaneous prediction. L1 generates a symbol to for next CC, not shown in the figure. The TPO's periodicity aligns with the paragraph's length. The cortex started in a basic state when the system started. At this point, all cortical columns are stable, so no CA3 is needed.

only when forward prediction of the next cortical column is active.

### Reinforced and Supervised Learning Support

Corticostriatal loops, through the matrix of the thalamus, can control the output of any cortical column in the cortex. This control has two effects: forcing EOS at will or shaping L1 inhibition results. This is done when the cortical column is in *unknown* state. In this way it will be possible to supervise the learning outcome by imprinting in L6b and L1 synapses. The guidance will be based on a replay mechanism (i.e., internal simulation of previously acquired knowledge). Sequence replay is selected according to rewarding outcomes. The corticostriatal loop may have certain effects helping L1 outcome result even during non-learning periods.

### Cortical Input Adaptation: The Case of the Auditory Pathway

HER (6) was specialized for speech processing, so the cortical input was conveniently adapted from the raw input specification. Only the following components in the auditory pathway were considered.

### Cochlea

It is a binary device that produces either one or zero in each band of frequency when the sound pressure is above or below a certain threshold. The threshold is controlled by the autonomic component of the descending auditory pathway. In particular, the posterior ventral cochlear nuclei set the threshold according to a predefined target of activity in the most recent ~25 seconds. The cochlear output has a high frequency (~1ms) to compensate for imprecise values.

### Dorsal Cochlear Nuclei

This component is a projector, like L1 and L4. It maps the cochlear output onto a reduced set of symbols. Beyond the output of the cochlea, cortical expectations (via L5b descending predictions, if the attention mechanism allows it) will contribute to the inhibition outcome. While the cochlear influence is fixed (i.e., there is no learning), cortical effects are learned in the cerebellar-like structures of the DCN. Thus, L5b predictions from the cortex are used to stabilize DCN output. The so-called embodiment requirement is just a stabilization of the input.

### Medial geniculate nucleus

Using a mechanism analogous to the HO nuclei, the MGN will filter out unexpected inputs from the auditory cortex. It will also suppress, via the glomerulus, fixed inputs to the cortex. This contrast-adaptation-like mechanism increases the information carried by the cortical input.

## Discussion

### Simulation based Evaluation.

The system as described was evaluated on (6), through the simulation of a synthetic system. A snapshot of how the system is able to process raw audio signal results is provided in Figure 2.

### Experimental Evidence Explanations.

Next, we provide some examples of possible explanations that the proposed model seems to offer.

- Being able to identify the sequence using only the last predicted value in L23/L5a is consistent with the "peak-end-rule" (14).
- Provides a plausible explanation for event segmentation(15) .
- The sequence segmentation observed in (6) is consistent with activity spikes with experimental evidence such as (16–19). Apparently, oscillations and bands in the EEG may be an epiphenomenon of sequence segmentation across the hierarchy.
- Dysfunction of the hippocampal complex explains both anterograde and retrograde amnesia. If it is not available, it could prevent the cortical columns from learning the new data due to the lack of long-term modulation (20). When it is incorrectly enabled it may induce catastrophic forgetting (21).
- Certain event-related potentials, such as the N400 or P600, may be explained by forward prediction or lack thereof (22).
- Top-down attention mechanism explains coherence changes in EEG after attentional engagement. (23).
- Forward prediction renders a large fraction of sensory input irrelevant, which may explain many attention-related phenomena such as the cocktail effect or many visual effects. An incorrect function of the mechanism in top-levels of the hierarchy may explain certain pathological conditions (24,25).
- Lateral contextualization may resolve the perception binding problem (26) and explain related effects such as McGurk's (27).

## Conclusions

This model may be misdirected, but its compactness, the synthetic performance, and its apparent ability to explain some of the most long-standing open issues of current neuroscience, encourage us to take the risk to further explore it. At best, it is likely to be only partially correct, but it provides an alternative path to follow. To the best of our knowledge, it is groundbreaking. Even if correct, it is obviously incomplete. The relevance of the brainstem is largely unexplored. The corticostriatal loop and the pre-mammalian structures attached to it, such as the hypothalamus, are missing. Movement, including the volitional sensor, will require this loop to function and is therefore also missing.

Nevertheless, the computer-brain analogy used earlier, from the discovery of the stored-program concept to the development of a modern computer system, will require considerable effort. However, without such a discovery, it may be impossible to understand how the remaining layers above it work. Nevertheless, one could argue that the distance between this early model and a competing system to a biological counterpart could be much smaller. The reliability of electronic system is much higher compared to a biological system, which may allow the elimination of many "engineering" complexities used in the latter.

This work provides a wealth of insights that challenge the most accepted models in neuroscience. While we were unable to find clear contradictions with the experimental evidence, we are excited to continue our research in this area. It is very likely that they do exist, so the model will need corrections. In any case, it can be useful as an alternative viewpoint to the reductionist impulse that neuroscience has followed. As F. Crick envisioned 30 years ago (28), in order to understand the brain, it will be necessary to broaden the focus by raising the level of abstraction.

## Acknowledgements

To José Angel Gregorio for the long, heated and productive discussions. Without them, this work would not have taken place.